\documentclass[12pt]{article}
\usepackage[dvips]{graphicx}
\usepackage{epsfig,cite}
\usepackage {amsmath}
\usepackage{color}
\usepackage{amssymb}
\usepackage{slashed}
\include{epsf}
\newcount\timecount
\newcount\hours \newcount\minutes  \newcount\temp \newcount\pmhours
\hours = \time
\divide\hours by 60
\temp = \hours
\multiply\temp by 60
\minutes = \time
\advance\minutes by -\temp
\def\hour{\the\hours}
\def\minute{\ifnum\minutes<10 0\the\minutes
            \else\the\minutes\fi}
\def\clock{
\ifnum\hours=0 12:\minute\ AM
\else\ifnum\hours<12 \hour:\minute\ AM
      \else\ifnum\hours=12 12:\minute\ PM
            \else\ifnum\hours>12
                 \pmhours=\hours
                 \advance\pmhours by -12
                 \the\pmhours:\minute\ PM
                 \fi
            \fi
      \fi
\fi
}

\def\monthname{\relax\ifcase\month 0/\or January\or February\or
   March\or April\or May\or June\or July\or August\or September\or
   October\or November\or December\else\number\month/\fi}

\def\bold#1{\setbox0=\hbox{$#1$}%
     \kern-.025em\copy0\kern-\wd0
     \kern.05em\copy0\kern-\wd0
     \kern-.025em\raise.0433em\box0 }

\def\beq{\begin{equation}}
\def\eeq{\end{equation}}

\def\ga{\mathrel{\raise.3ex\hbox{$>$\kern-.75em\lower1ex\hbox{$\sim$}}}}
\def\la{\mathrel{\raise.3ex\hbox{$<$\kern-.75em\lower1ex\hbox{$\sim$}}}}
\def\gev{{\rm \, Ge\kern-0.125em V}}
\def\tev{{\rm \, Te\kern-0.125em V}}
\def\gyr{{\rm \, G\kern-0.125em yr}}

\def\gappeq{\mathrel{\rlap {\raise.5ex\hbox{$>$}}
{\lower.5ex\hbox{$\sim$}}}}
\def\lappeq{\mathrel{\rlap{\raise.5ex\hbox{$<$}}
{\lower.5ex\hbox{$\sim$}}}}
\def\Toprel#1\over#2{\mathrel{\mathop{#2}\limits^{#1}}}

\def\m12{m_{1\!/2}}

\newcommand\iso[2]{\mbox{${}^{#2}${\rm #1}}}
\def\he#1{\iso{He}{#1}}

\def\li#1{\iso{Li}{#1}}

\def\bea{\begin{eqnarray}}
\def\eea{\end{eqnarray}}

\def\nrg{\epsilon}

\def\pfrac#1#2{\left(\frac{#1}{#2}\right)}
\def\avg#1{\langle #1 \rangle}
\def\beqar{\begin{eqnarray}}
\def\eeqar{\end{eqnarray}}

\def\gann{\Gamma_{\rm ann}}
\def\sigva{\avg{\sigma v}_{\rm ann}}

\def\nrg{\epsilon}

\def\pfrac#1#2{\left( \frac{#1}{#2} \right)}

\begin{document}

\begin{titlepage}
\pagestyle{empty}
\baselineskip=21pt
\rightline{KCL-PH-TH/2011-26, LCTS/2011-12, CERN-PH-TH/2011-205}
\rightline{UMN--TH--3009/11, FTPI--MINN--11/19}
\vskip 0.1in
\begin{center}
{\large {\bf An Enhanced Cosmological  $\bf  {}^{6} Li$ Abundance as a Potential Signature of Residual Dark Matter Annihilations}}

\end{center}
\begin{center}
 {\bf John~Ellis}$^{1,2}$,
 {\bf Brian~D.~Fields}$^3$,
{\bf Feng~Luo}$^4$,
{\bf Keith~A.~Olive}$^{4,5}$
and {\bf Vassilis~C.~Spanos}$^6$
\vskip 0.1in
{\small {\it
$^1${Theoretical Particle Physics and Cosmology Group, Physics Department, \\
King's College London, London WC2R 2LS, UK}\\
$^2${TH Division, Physics Department, CERN, CH-1211 Geneva 23, Switzerland}\\
$^3${Center for Theoretical Astrophysics, Departments of Astronomy and of Physics, \\
University of Illinois, Urbana, IL 61801, USA}\\
$^4${School of Physics and Astronomy, \\
University of Minnesota, Minneapolis, MN 55455, USA}\\
$^5${William I. Fine Theoretical Physics Institute, \\
University of Minnesota, Minneapolis, MN 55455, USA}\\
$^6${Institute of Nuclear Physics, NCSR ``Demokritos'', GR-15310 Athens, Greece}} \\
}
\vskip 0.2in
{\bf Abstract}
\end{center}
\baselineskip=18pt \noindent
{\small

Residual late-time dark matter particle annihilations during and after Big-Bang
Nucleosynthesis (BBN) may alter the predicted cosmological abundances of the light
elements. Within the constrained minimal supersymmetric extension of the
Standard Model (the CMSSM) with a neutralino LSP, we find negligible effects
on the abundances of Deuterium, \he3, \he4 and \li7 predicted by homogeneous BBN, but potentially a large
enhancement in the predicted abundance of \li6. This enhancement may be as much as
two orders of magnitude in the focus-point WMAP strip 
and in the coannihilation and funnel regions for large $\tan \beta$ for small $m_{1/2}$, and the effect is still significant
at large $m_{1/2}$. However, the potential \li6
enhancement is negligible in the part of the coannihilation strip for $\tan \beta = 10$
that survives the latest LHC constraints. A similar enhancement of the \li6 abundance
may also be found in a model with common, non-universal Higgs masses (the NUHM1).}


\vfill
\leftline{
September  2011}
\end{titlepage}
\baselineskip=18pt

\section{Introduction}

The success of homogeneous Big-Bang Nucleosynthesis (BBN) is one of the
lynchpins of cosmology. Using the baryon-to-photon ratio, $\eta$, inferred from
measurements of the cosmic microwave background (CMB) radiation \cite{wmap7},
homogeneous BBN predicts successfully the astrophysical abundances of 
Deuterium, \he3 and \he4~\cite{cfo1,cfo2,bbn2,fs,cyburt,cfo5}. On the other hand, there are issues with the
abundances of \li7 \cite{cfo5} and potentially with \li6 \cite{rvo}. 
In particular, the predicted abundance of \li7 is
considerably larger than the range suggested by 
observations~\cite{spite,rbn,rbofn,liglob,liglob2,asp06,hos,aoki,sbordone,newer}~\footnote{A
globular cluster star with a \li7 abundance comparable to the BBN prediction has recently been observed~\cite{LiNa}:
this value may be due to production by a previous generation of stars.}, and there
are suggestions that the astrophysical value of the \li6 abundance may be
much higher than predicted by homogeneous BBN \cite{asp06}. However, one should note that  
the line asymmetries which have been interpreted as \li6 could be the result of
convective processes affecting \li7 \cite{cayrel}.

We and others have investigated previously whether the late decays of massive particles,
such as the gravitino in the constrained minimal supersymmetric extension of the
Standard Model (the CMSSM) with a neutralino as the lightest supersymmetric particle (LSP), 
could improve significantly the \li6 and \li7
abundances predicted by homogeneous BBN~\cite{Lindley:1984bg,Ellis:1984er,Lindley:1986wt,Scherrer:1987rr,Reno,Dimopoulos:1988ue,ellis,Kawasaki:1994af,kawmoroi,holtmann,karsten,kkm,kohri,cefo,jed,Jedamzik:2004ip,kkm2,EOV,kmy,Jed1,Jed2,Jedamzik06,stef,jp,ceflos}. We did not find a solution to the \li6 
problem, but we did find a region of supersymmetric parameter space where
gravitino decays might alleviate or even solve the \li7 problem~\cite{ceflos,ceflos2}. On the other hand
the \li7 problem might have a more banal solution, such as the existence of a
suitable carbon, boron or beryllium resonance~\cite{resonance}.

In addition to decays, the late-time annihilations of cold dark matter may also affect the abundances of the light elements
\cite{Reno,earlylate,hisano,Jedamzik:2004ip,jed}.
In particular,  these annihilations may have a significant 
effect on the abundance of \li6  \cite{Jedamzik:2004ip}.  There it was argued that \li6 production may occur
if the $s$-wave annihilation cross-section is sufficiently large, and it was assumed that the relic density
of the annihilating dark matter particles is controlled largely by the $s$-wave part of the cross-section.
However, in supersymmetric models where the LSP is a neutralino, such as the CMSSM,
the relic density is in fact largely determined by the
$p$-wave part of the cross-section, which by the time of BBN is essentially ineffective. Therefore,
a re-analysis of the suggestion of~\cite{Jedamzik:2004ip} in the context of the CMSSM and related
models is timely, and is the subject of this paper.

In this paper we study the possible effects on the cosmological light-element
abundances of residual late-time annihilations of neutralino LSPs during or after BBN~\cite{earlylate,jed,Jedamzik:2004ip}.\footnote
{
Recent papers have also considered the BBN consequences of
WIMP models having residual annihilations increased by
Sommerfeld or Briet-Wigner enhancements \cite{enhance}.
In the CMSSM model, neither effect occurs,
due to the lack of a light boson and of 
extreme degeneracy in the funnel, respectively.
}
We find negligible effects on the abundances of Deuterium, \he3, \he4 and \li7 predicted 
by homogeneous BBN, but potentially a large enhancement in the predicted abundance of \li6,
as suggested in~\cite{Jedamzik:2004ip}.
The physics of this effect is the following. It is well understood that the famous $A = 5$ gap
in the spectrum of stable nuclei impedes the production of heavier nuclei in BBN. 
The dominant mechanism for making \li6 in annihilating-particle scenarios
is initiated by $p$ and $n$ spallation of \he4.  This yields many $A = 3$ nuclei with only a tiny
reduction in \he4 abundance. The tritium and \he3 nuclei are produced with
large, nonthermal energies, and subsequently slow down due to 
ionization losses,
but have some probability of inducing $t (\alpha, n)\li6$ or $\he3 ( \alpha, p) \li6$
reactions first. In this way, an amount of \li6 may be produced that is large relative
to the standard homogeneous BBN abundance, without making large
amounts of extra Deuterium and $A = 3$ or reducing the \he4 abundance, and leaving the \li7
abundance unaffected.

\begin{figure}[htb]
\begin{center}
\epsfig{file=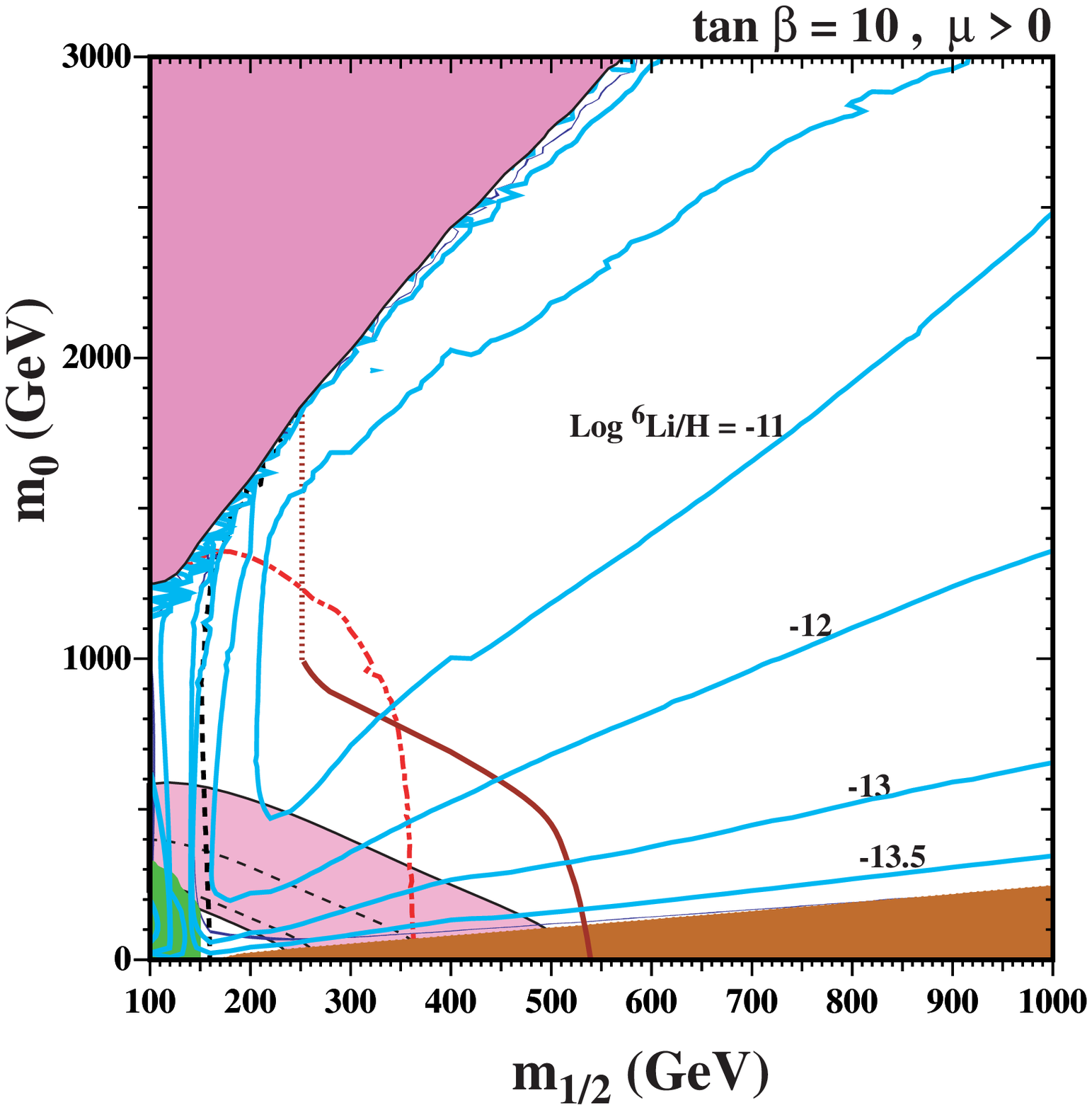,width=0.475\textwidth}
\epsfig{file=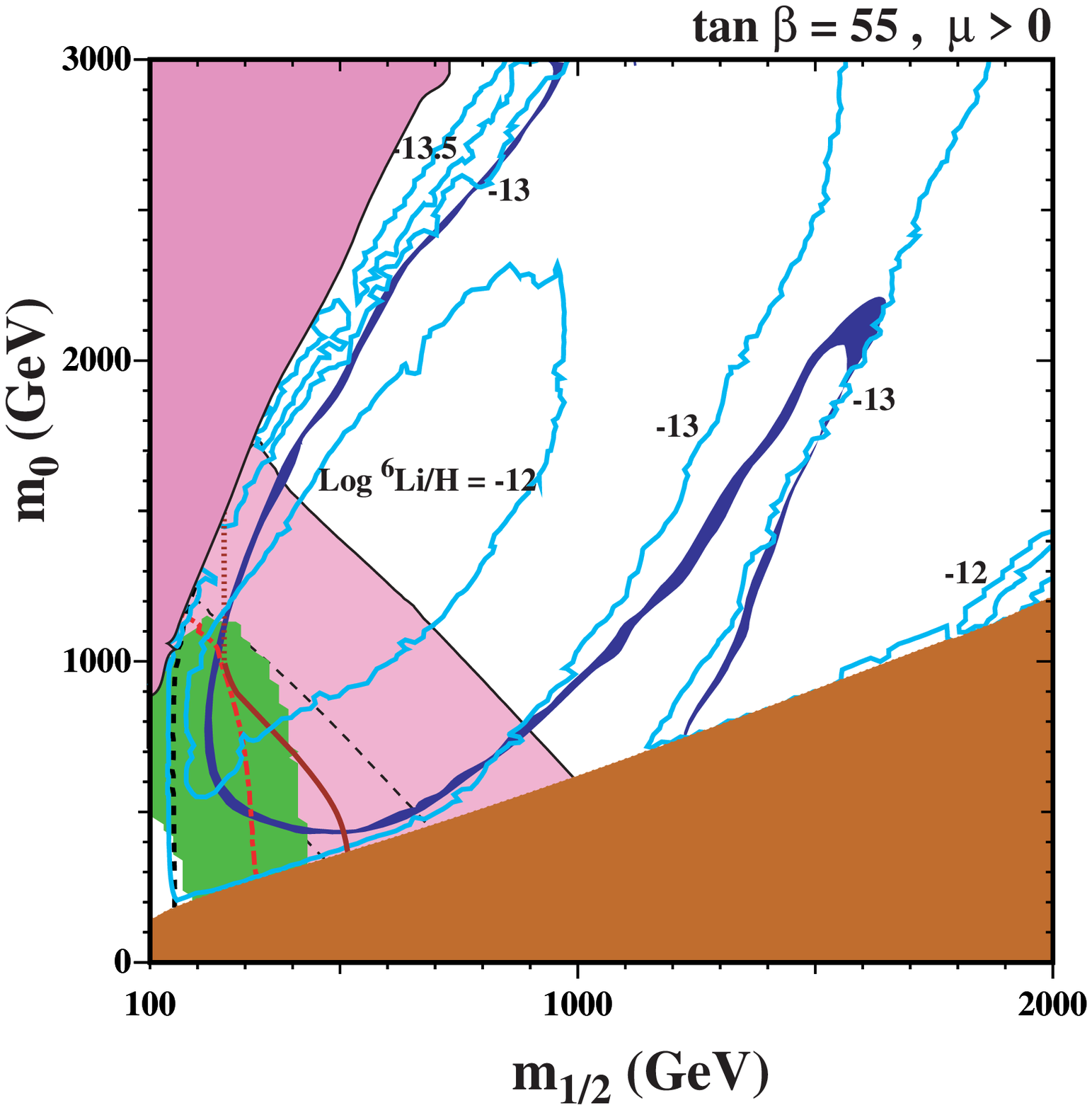,width=0.48\textwidth}
\caption{
\it Left: The CMSSM $(m_{1/2}, m_0)$ plane for $A_0 = 0$ and $\tan \beta = 10$,
and Right: the corresponding plane for $\tan \beta =55$, both with $\mu > 0$,
displaying contours of the \li6 abundance including the effects of late-time
$\chi \chi$ annihilations. Contours of the \li6 abundance
are coloured light blue, and the WMAP-compatible~\protect\cite{wmap7} strips of
parameter space are shaded dark blue. The brown shaded region at large $m_{1/2}$ and 
small $m_0$ is excluded because the LSP would be charged, and in the pink
shaded region at small $m_{1/2}$ and large $m_0$ there would be no consistent
electroweak vacuum. Also shown are the exclusion by LEP searches
for the Higgs boson (red dash-dotted line) and charginos (black dashed line), 
and by LHC searches for sparticles (purple solid line)~\protect\cite{LHC,CMS}. 
The green shaded region is excluded by $b \to s \gamma$, and the paler pink region is favoured by
$g_\mu - 2$ at the 1- (2-)$\sigma$ level, as indicated by the dashed (solid) black lines.}
\label{fig:planes}
\end{center}
\end{figure}

In the CMSSM~\cite{cmssm1}, it is assumed that all supersymmetry-breaking gaugino masses have a common value
$m_{1/2}$ at some grand unification scale before renormalization, and likewise all the soft
supersymmetry-breaking scalar masses are assumed to have a common $m_0$. The other
parameters of this model are the (supposedly) universal trilinear parameter $A_0$  (taken here to be $A_0 = 0$),
and the ratio of Higgs vacuum expectation values, $\tan \beta$. In addition, one must specify the 
sign of the Higgs mixing mass, $\mu$, which is generally taken to be positive in the CMSSM so as to
improve compatibility with measurements of $g_\mu - 2$ and $b \to s \gamma$ decays. 
As is well known, in the CMSSM there are strips in the $(m_{1/2}, m_0)$ planes
for fixed $A_0$ and $\tan \beta$ 
along which the relic density of the
neutralino LSP, $\chi$, lies within the range favoured by WMAP and other
astrophysical observations~\cite{cmssmwmap}. At relatively low values of $m_0$ there is generically
a coannihilation strip close the boundary where the LSP would become charged
that extends, for large $\tan \beta$, into a funnel at large $m_{1/2}$ where the LSPs annihilate rapidly
via direct-channel heavy Higgs resonances.
At relatively high values of $m_0$, there is a focus-point strip close to the boundary of
consistent electroweak symmetry breaking. In light of present experimental constraints from
the LHC and elsewhere, plausible values of $\tan \beta$ range between $\sim 10$ and $\sim 55$~\cite{LHC}.
Representative $(m_{1/2}, m_0)$ planes for these values of $\tan \beta$ are shown in Fig.~\ref{fig:planes},
and are discussed below in more detail.
In this paper, we explore the effects on cosmological light-element abundances of residual 
late-time $\chi \chi$ annihilations for CMSSM parameters along the coannihilation/funnel
and focus-point strips for these reference values of $\tan \beta$, and also remark on additional
possibilities in one- and two-parameter generalizations of the CMSSM with non-universal soft supersymmetry-breaking contributions to
Higgs masses (the NUHM1 and NUHM2)~\cite{nuhm1,nuhm12,nuhm2}.

We find that such late annihilations have no significant effect on the cosmological
abundances of Deuterium, \he3, \he4 and \li7 in any of the CMSSM scenarios studied.
However, they may enhance the \li6 abundance by up to two orders of magnitude: in some
instances, we find \li6/H $\sim 10^{-12}$, compared to the value $\sim 10^{-14}$ found in
standard homogeneous BBN \cite{bbnli6,van}. This possibility arises at low $m_{1/2}$ along the
focus-point strips for $\tan \beta = 10$ and 55, and also along the coannihilation strip
for $\tan \beta = 55$. The values of \li6/H along these strips decrease to $\sim 10^{-13}$
at large $m_{1/2}$, which is typical also of values in the funnel region for $\tan \beta = 55$.
On the other hand, we find no substantial enhancement of \li6/H along the
coannihilation strip for $\tan \beta = 10$ for values of $m_{1/2}$ consistent with the LHC
constraints~\cite{LHC,CMS}, though values as large as $\sim 10^{-13}$ might have been reached at
lower $m_{1/2}$. The possibility of a large \li6 enhancement is extended in the NUHM1
to a large range of $m_0$ with $\tan \beta = 10$ and low $m_{1/2}$.

\section{The Cosmological Lithium Problems}

Cosmologically, the dominant Lithium isotope is \li7, whose abundance is commonly
inferred from observations of low-metallicity halo dwarf stars. These indicate a plateau of
Lithium versus metallicity~\cite{spite}, with 
\begin{equation}
\left( \frac{\li7}{\rm H} \right)_{\rm halo*} \; = \; (1.23^{+0.34}_{-0.16}) \times 10^{-10} ,
\label{Li7Hhalo}
\end{equation}
whereas observations of globular clusters~\cite{liglob} yield somewhat higher values:
\begin{equation}
\left( \frac{\li7}{\rm H} \right)_{\rm GC} \; = \; (2.35 \pm 0.05) \times 10^{-10} .
\label{Li7HGC}
\end{equation}
For comparison, the standard BBN result for \li7/H is $(5.12^{+0.71}_{-0.62}) \times 10^{- 10}$~\cite{cfo5}, 
and the difference between this and (\ref{Li7Hhalo}, \ref{Li7HGC}) constitutes the
cosmological \li7 problem. As already mentioned, this might be resolved by new
physics beyond the Standard Model such as late decays of massive gravitinos~\cite{ceflos2},
or by some undocumented Standard Model effect such as a suitable carbon, boron or
beryllium resonance~\cite{resonance}. The cosmological \li7 problem is not our focus in this paper.

\li6 has been observed in some halo stars \cite{li6obs} with [Fe/H] $\sim$ -2, and
with an isotopic ratio that is 
\begin{equation}
\left( \frac{\li6}{\li7} \right)_{\rm halo*} \; \sim \; 0.05 .
\label{Li6Li7}
\end{equation}
These observations are consistent with the results of Galactic cosmic-ray (GCR) nucleosynthesis \cite{gcr,van,fo}, though see below for results
at lower metallicity.
This confirms that most of the Lithium is in the form of \li7, leaving unscathed the cosmological
\li7 problem.

However, a recent paper has reported the presence of a similar isotopic abundance in halo stars
over a broad range of metallicities that extends to significantly lower
values ([Fe/H] $\sim$ -1 to -3)  \cite{li6obs,asp06}.
The inferred plateau \li6/H ratio $\sim (6$ to $25) \times 10^{-12}$
is about 1000 times higher than the \li6/H ratio predicted by standard homogeneous BBN~\cite{bbnli6,van},
namely $\li6/{\rm H} \sim 10^{-14}$. The isotopic ratio (\ref{Li6Li7}) 
cannot be explained by conventional GCR nucleosynthesis, 
at the lowest metallicities: 
this is the cosmological \li6 problem. The reliability of the \li6 plateau at very
low metallicity has been questioned \cite{cayrel},
so the \li6 problem should be taken with a grain of salt. But in any
case, these exiting if controversial results demonstrate that 
\li6 abundances at levels $\li6/{\rm H} \la  {\rm few} \times 10^{-12}$
are at or near the reach of present observational techniques.

Thus the current observational situation is evolving, but 
without question is interesting:  
at the very least, the present results serve as upper limits to
primordial \li6, 
and impose bounds on nonstandard BBN.
At most, current data may already point to a primordial
\li6 problem which would {\em demand} new BBN physics,
and probe its details.
Our focus in this paper is to determine the \li6 production
and its observational implications in the context of
some of the most popular supersymmetric dark matter scenarios.

It has been proposed that some decaying-particle scenario might produce \li6 at the plateau level 
with some destruction of \li7~\cite{Jed2,cefos,grant,jed3,cumber,bailly}, 
offering the possibility of solving both Lithium problems simultaneously.
However, we note that solving the \li6 problem would use up only a small fraction of the \li7 whose
destruction would be needed to solve the \li7 problem, leading one to consider separate solutions
for the two Lithium problems. It is also possible that the \li6 problem might be explained
by nucleosynthesis due to cosmological cosmic rays produced at the epoch of structure formation~\cite{rvo,ccr}.
We have previously demonstrated that late-decaying massive gravitinos
might resolve the \li7 problem within the CMSSM framework~\cite{ceflos2}. Here we show that the \li6 problem
might, independently and in parallel, have at least a partial supersymmetric solution, 
via the late annihilations of neutralino LSPs.

\section{Residual Late-Time Neutralino Annihilations}

Assuming that the lightest neutralino $\chi$ is the LSP, and that R-parity is conserved,
the relic neutralino density is essentially fixed at a freeze-out temperature $T_f \sim m_\chi/20$.
At lower temperatures, the local density of neutralinos, $n_\chi$, decreases as  the universe expands
(presumably) adiabatically, and subsequent annihilations have very little effect on the dark matter
density, but may have important effects on the light-element abundances~\cite{Reno,earlylate,Jedamzik:2004ip,jed}.

The rate per volume of annihilation {\em events} is
\beq
q_{\rm ann} = \frac{1}{2} n_\chi^2 \sigva ,
\label{qrate}
\eeq
and so the annihilation event rate {\em per $\chi$} is
\beq
\gann = \frac{q_{\rm ann}}{n_{\chi}} 
 = \frac{1}{2}\sigva n_\chi = \frac{1}{2}\sigva Y_\chi n_{\rm b} ,
\eeq
and thus the annihilation event rate {\em per baryon} is 
\beq
\label{eq:decaysperb}
\frac{q_{\rm ann}}{n_{\rm b}} 
= \gann Y_\chi
= \frac{1}{2}\sigva Y_\chi^2 n_{\rm b} ,
\eeq
where the $\chi$ abundance is
\beq
Y_\chi = \frac{n_\chi}{n_{\rm b}} 
 = \frac{m_{\rm b}}{m_\chi} \frac{\Omega_\chi}{\Omega_{\rm b}} \, .
\eeq
At the temperatures of interest here, 
$T_{\rm BBN} \la 1 \ {\rm MeV} \ll m_{\chi}$,
the annihilation rate coefficient $\sigva$ is very well approximated
as a constant, the value of which depends on the specific underlying
supersymmetry model. In (\ref{qrate}), we are interested in only the $s$-wave part of the 
cross-section whereas a combination of $s$- and (mainly) $p$-wave cross-sections is constrained by the requirement of 
reproducing the present dark matter density
within errors.

The annihilations inject nonthermal 
Standard Model particles, including both 
electromagnetic as well as hadronic species.
For electromagnetic products we need only track the
total energy injected per annihilation.
For nonthermal hadrons (nucleons)
$h = n, p$, we calculate the spectrum
$Q_h^{\rm ann}(\nrg)$ of annihilation products, 
normalized such that 
$\int Q_h^{\rm ann} (\nrg) \ d\nrg = B_h$,
the expected number of $h$ created per annihilation.
Then the injection/source rate of $h$ due to annihilations,
per unit volume, per unit time, and per unit kinetic energy $\nrg$, is
\beq
\frac{d{\cal N}_{h,\rm inj}^{\rm ann}}{dV \, dt \, d\nrg}
 = q_{\rm ann} Q_h^{\rm ann}(\nrg)  \  .
\eeq

These particles then lose energy as they propagate
in the cosmic plasma.  The propagated spectrum of nonthermal
particles must be calculated, and this
produces the nonthermal reactions on ambient thermal light nuclides
that perturb BBN.

The effect of nonthermal particle injection in BBN has
been well-studied in the case of decays of some unstable
particle $X$.  Much of the physics
carries over here, once one makes the appropriate substitution
of abundances $n_X \rightarrow n_\chi$ and
of annihilation rate for decay rate:
$\Gamma_X  = \tau_X^{-1} \longrightarrow \gann$.
After injection, the nonthermal particle propagation remains the
same, and we treat this as in \cite{ceflos}.
We also adopt the same set of nonthermal BBN reactions
which we include in the same manner, making the
appropriate substitution of annihilations for decays.

\subsection{Order-of-Magnitude Calculation}

Before turning to our numerical results,
we first present an order-of-magnitude calculation that illustrates
the basic physics in play, and also serves as a check on our numerical results.
The total number of annihilation events
per baryon occurring after a given time $t_i$  is
the time integral of eq.~(\ref{eq:decaysperb})
\beqar
{\cal N}_{\rm ann} & = & \int_{t_i} \gann \ Y_\chi \ dt
  \sim Y_\chi^2 \sigva n_{\rm b}(t_i) \ t_i  \\
  & = & 5 \times 10^{-9} \ {\rm events/baryon} \
    \pfrac{\sigva}{10^{-26} \ \rm cm^3 \, s^{-1}} \
    \pfrac{300 \rm GeV}{m_\chi}^2 \ \ .
\eeqar
Our fiducial values correspond to $t_i \sim 100 \ \rm sec$ 
and $T_i \sim 100 \ \rm keV$,
since this marks the epoch when the \he4 
abundance becomes large.

Given this number of annihilations per baryon, we now need
the branching for \li6 production per annihilation.
As discussed earlier, nonthermal particles from annihilations or decays produce \li6 
as a secondary by-product of \he4 spallation:
\beqar
\nonumber
p_{\rm nonthermal} \he4 & \rightarrow & \iso{H}{3}_{\rm nonthermal}  + \cdots \\
 & &  \iso{H}{3}_{\rm nonthermal} + \he4 \rightarrow \li6 + n  \, ,
\eeqar
and similarly with nonthermal $\he3$; nonthermal D also contributes but
is subdominant.
As discussed for the late-decay case in~\cite{ceflos}, each late annihilation produces a mass-3
abundance increment $\Delta Y({}^{3}A)$ which is given in the thin target limit by
\beq
\Delta Y({}^{3}A) \sim {\cal N}_{\rm ann} B_N
    \frac{\sigma(N\alpha \rightarrow {}^{3}A+\cdots)}
         {\sigma(N\alpha \rightarrow {\rm inelastic})} ,
\eeq
where $B_N \sim 0.4$ is the number of nucleons per annihilation.
Typically this increases the mass-3 abundance by an amount
$\Delta Y({}^3A) \sim 10^{-9} \ll Y_{\rm BBN}({}^3A) \sim 10^{-5}$, i.e.,
much smaller than the standard primordial abundance,
and thus we do not expect substantial perturbations to mass-3 nuclides, or to D,
which has similar cross sections, or to \he4.

The energetic $A=3$ particles are slowed
in the cosmic plasma by ionization and related losses,
with a range $R_3 = \int (dE/dX)^{-1} dE$, where
$dE/dX$ is the loss rate per
thickness $dX = \rho_{\rm b} dx$ in $[\rm g/cm^2]$. 
Hence the stopping length is $R_3/\rho_{\rm b}$.
The fraction of mass-3 nuclides which produce \li6
before stopping is this stopping length
divided by the mean free path for \li6 production, namely:
\beq
f({}^{3}A \rightarrow \li6) \sim  n_\alpha \sigma({}^{3}{A}\alpha \rightarrow \li6) \frac{R_3}{\rho_{\rm b}}
  \sim Y_\alpha \frac{\sigma({}^{3}{A}\alpha \rightarrow \li6) R_3}{m_{\rm b}}
  \sim 7 \times 10^{-4} \ \ .
\eeq
Collecting these results, the residual late-time annihilation
contribution to the \li6 abundance per baryon is
\beqar
\Delta Y(\li6) & = &  \Delta Y({}^{3}A) 
   \ f({}^{3}A \rightarrow \li6)  \\
  & \sim &  B_N  Y_\chi^2 \sigva Y_\alpha 
    \frac{\sigma(N\alpha \rightarrow {}^{3}A+\cdots)}
        {\sigma(N\alpha \rightarrow {\rm inelastic})} \
     \frac{\sigma({}^{3}{A}\alpha \rightarrow \li6) R_3}{m_{\rm b}} \
  n_{\rm b}(t_i) \ t_i \\
 & = & 7 \times 10^{-13} \,
    \pfrac{\sigva}{10^{-26} \ \rm cm^3 \, s^{-1}} \,
    \pfrac{300 \ \rm GeV}{m_\chi}^2 \ \ .
    \label{FoM}
\eeqar

The numerical results given above are evaluated for $t_i = 100 \ \rm sec$,
and we also take $R_3 = 1 \, \rm g/cm^2$ and
$\sigma({}^{3}{A}\alpha \rightarrow \li6) = 30 \ \rm mb$.
This formula gives the scaling 
$\Delta Y(\li6) \propto B_N \sigva (\Omega_\chi/m_\chi)^2$
which we verify with our full numerical results,
and the normalization agrees to within a factor $\sim 2$.
This agreement lends confidence in our code and our 
understanding of the physics.

\subsection{Numerical Results}

We turn now to our full numerical results. In order to establish the context for our
subsequent analysis of the possible annihilation effects along the strips in
CMSSM parameter space that are compatible with WMAP and other constraints
on the present-day dark matter density, we first discuss the full CMSSM
$(m_{1/2}, m_0)$ planes shown in Fig.~\ref{fig:planes}. The light blue lines
are contours of the \li6 abundance, and the relic density is WMAP-compatible~\cite{wmap7}
along the dark blue strips, assuming that the lightest neutralino $\chi$ is the LSP
and is stable, as in $R$-conserving models. There would be no consistent
electroweak vacuum in the pink shaded region at small $m_{1/2}$ and large $m_0$,
the lighter ${\tilde \tau}$ would be the LSP in the brown shaded region, and the
green shaded region is excluded by $b \to s \gamma$ decay~\footnote{According to
conventional Big-Bang cosmology and in the absence of $R$ violation,
the LSP $\chi$ would be overdense in the regions between the WMAP strips. It would be
underdense in the regions between these strips and the pink and brown shaded regions.}. Regions to the left of
the red dash-dotted (black dashed) (purple) line are excluded by searches
for the Higgs boson at LEP (charginos) (LHC searches for sparticles)~\cite{LHC,CMS}.
In the paler pink region the supersymmetric contribution remedies the discrepancy between the experimental measurement of
$g_\mu - 2$ and theoretical calculation within the Standard Model using low-energy $e^+ e^-$ data, 
with 1- (2-)$\sigma$ consistency being indicated by the dashed (solid) black lines.


We see in the left panel of Fig.~\ref{fig:planes} showing the $(m_{1/2}, m_0)$ plane for $\tan \beta = 10$ that
most of the lower (coannihilation) WMAP strip has \li6/H $< 10^{-13}$, whereas the upper
(focus-point) strip may have \li6/H as large as $10^{-12}$. There is a region where \li6/H
seems able to exceed $10^{-11}$, but this is well inside the region between the WMAP strips,
where the relic $\chi$ is overdense according to conventional Big-Bang cosmology. 
In the right panel of Fig.~\ref{fig:planes} for $\tan \beta = 55$,
we see that \li6/H $\sim 10^{-13}$ along the coannihilation strip and in the funnel region at large
$m_{1/2}$ and $m_0$ where the relic density is brought into the WMAP range by rapid
annihilations through direct-channel $H/A$ resonances, though somewhat larger values of
\li6/H are possible at small $m_{1/2}$. Along the focus-point strip, we see that values of
\li6/H $\sim 10^{-12}$ are also possible at small $m_{1/2}$, falling to $\sim 10^{-13}$
at large $m_{1/2}$. The range \li6/H $\sim 10^{-11}$ is never attained for $\tan \beta =55$,
even in the overdense region of the $(m_{1/2}, m_0)$ plane.

\begin{figure}[htb!]
\begin{center}
\epsfig{file=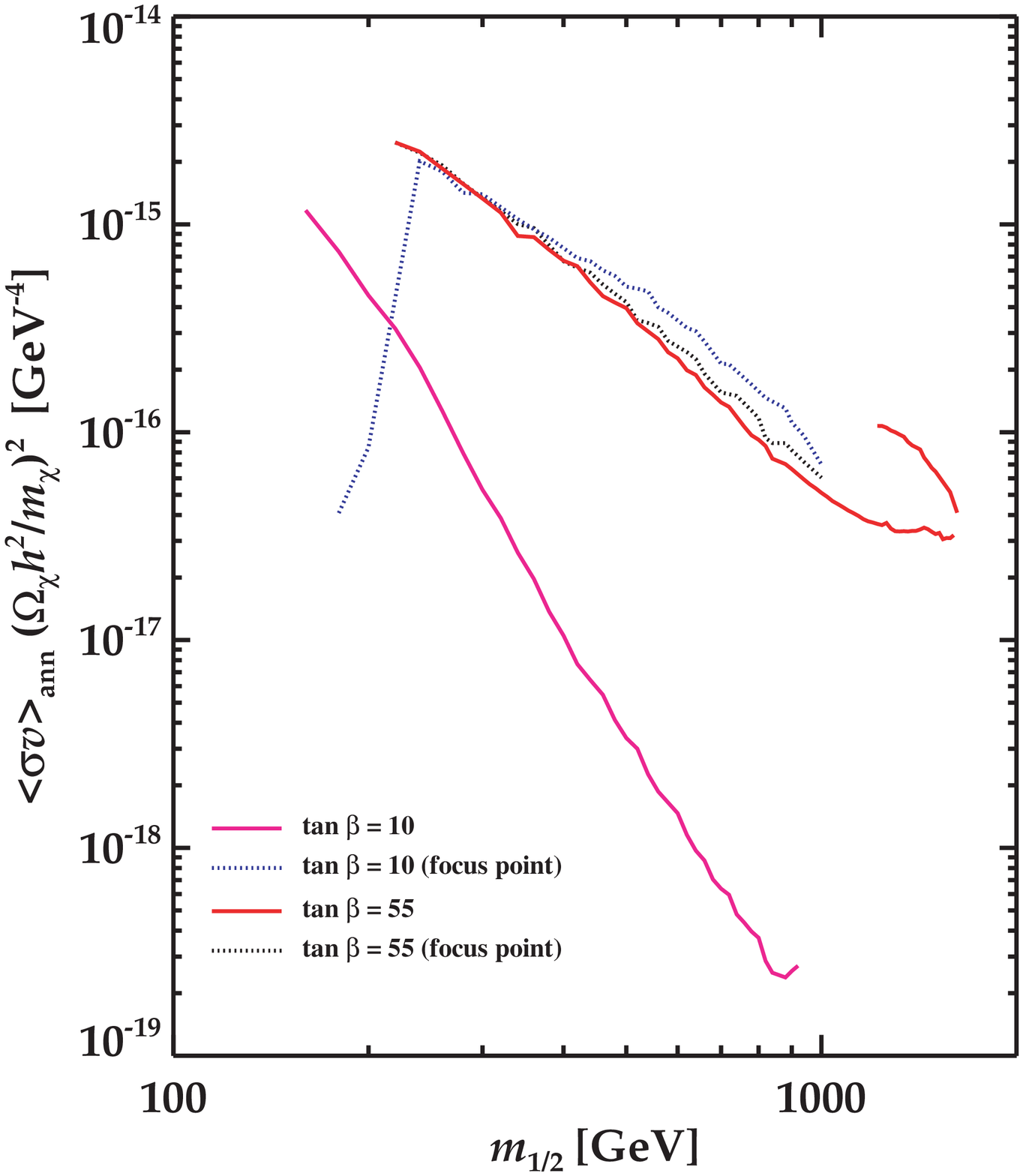,width=0.4775\textwidth}
\epsfig{file=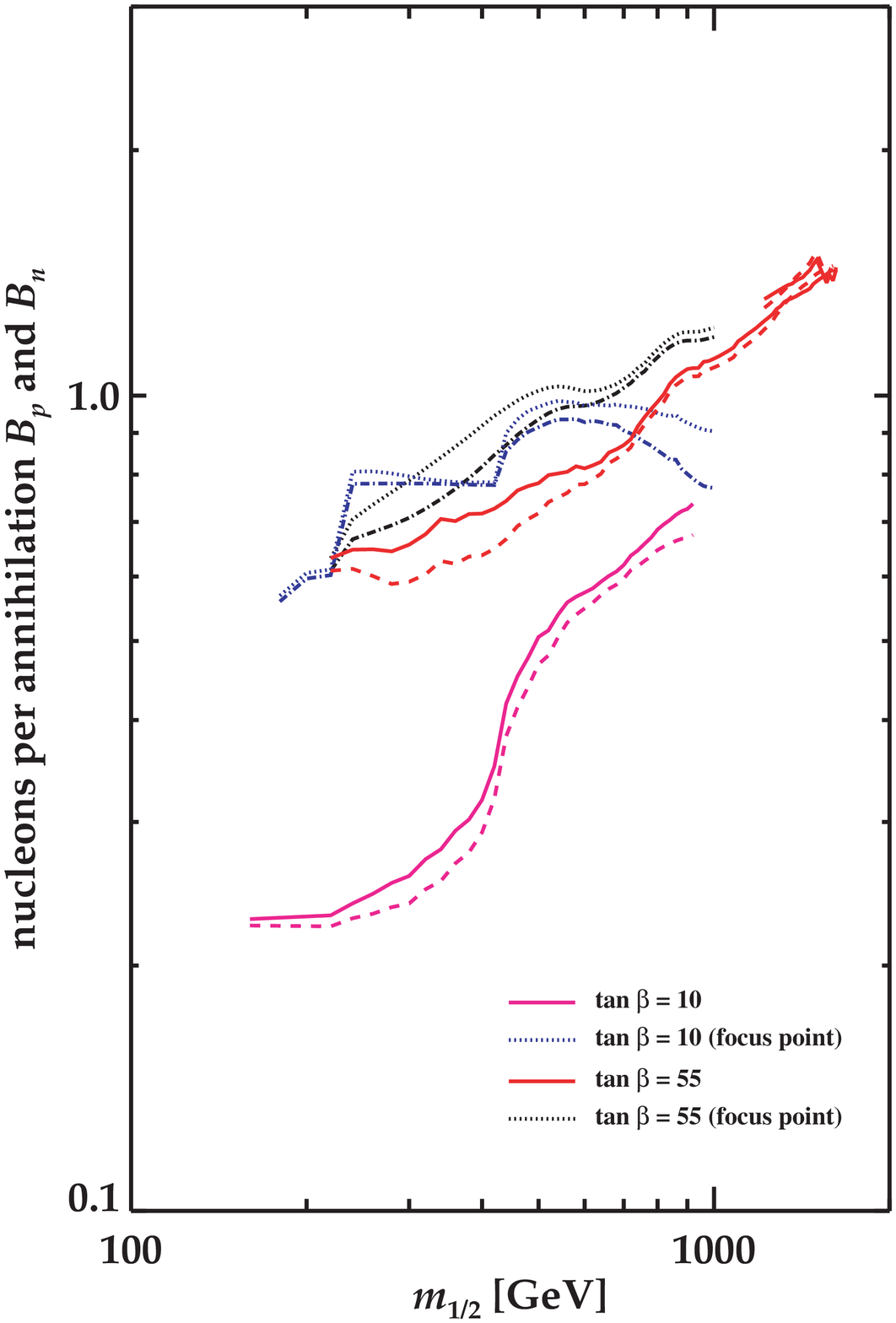,width=0.37\textwidth}
\caption{
\it Left: The figure of merit for the late-time annihilation rate, namely the velocity-averaged
$\chi  \chi$ annihilation cross section divided by the square of the neutralino mass,
$\sigva (\Omega_\chi h^2/m_\chi)^2$, along the WMAP strips in the
coannihilation, focus-point and funnel regions for $\tan \beta = 10, 55$,
$A_0 = 0$ and $\mu > 0$, as functions of $m_{1/2}$. We see that the figure of merit
along the $\tan \beta = 10$ coannihilation strip is much smaller
than along the other strips, and that all decrease rapidly as $m_{1/2}$ increases.
Right: The numbers $B_{p,n}$ of protons (solid or dotted lines) and neutrons (dashed or dash-dotted lines) produced per $\chi \chi$ annihilation
event, as calculated using {\tt PYTHIA}, along the WMAP strips in the
coannihilation, focus-point and funnel regions for $\tan \beta = 10, 55$,
$A_0 = 0$ and $\mu > 0$, as functions of $m_{1/2}$. We see that in general
the numbers increase significantly as $m_{1/2}$ increases.}
\label{fig:ratefom}
\end{center}
\end{figure}

We now focus on the WMAP strips in Fig.~\ref{fig:planes}. The left panel of
Fig.~\ref{fig:ratefom} displays the figure-of-merit combination
$\sigva (\Omega_\chi h^2/m_\chi)^2$ as a function of $m_{1/2}$ along the WMAP strips in the
CMSSM for $\tan \beta = 10$ and 55. 
We see that $\sigva (\Omega_\chi h^2/m_\chi)^2$ along the coannihilation
strip for $\tan \beta = 10$ is much smaller than along the other strips. This can be
understood from the fact that along this strip several coannihilation processes
involving sleptons contribute to reducing the relic $\chi$ density into the WMAP range,
and that their relative contributions become more important as $m_{1/2}$ increases.
In addition, along this strip the $s$-wave cross-section relevant during BBN
is significantly smaller than the $p$-wave cross-section that dominates during freeze-out.
These coannihilation processes are less important along the corresponding strip for
$\tan \beta = 55$, and unimportant in the funnel region and along the focus-point
strips, where the $s$-wave cross-section becomes 
comparable to the total cross section.
Hence, along these strips $\sigva$ must be larger, in
order to bring the relic density down into the WMAP range unaided.

The $\chi \chi$ annihilations feed many different particle species into the
cosmological background, initially with nonthermal spectra that we model using {\tt PYTHIA}~\cite{pythia}. 
The only species that survive long enough to interact significantly with background nuclei are
protons and neutrons (and their antiparticles) and photons. The former are far
more important for the nuclear reactions of interest here, so we focus on their
numbers and spectra. The right panel of Fig.~\ref{fig:ratefom} displays the numbers of protons (solid or dotted lines) and neutrons (dashed or dash-dotted lines) produced per annihilation event, again along the WMAP strips
for $\tan \beta = 10, 55$ discussed previously. We see that in general the numbers of 
protons and neutrons increase significantly as $m_{1/2}$ increases, with some bumps
as new annihilation thresholds are crossed.

Fig.~3 of~\cite{EOSgamma} displays the most important branching fractions for
final states in $\chi \chi$ annihilations as functions of $m_{1/2}$ along the
WMAP strips for $\tan \beta = 10$ and 55, which include the
final states $\tau^+ \tau^-$, $b {\bar b} $, $W^+ W^-$, $t {\bar t} $, $hZ$ and $ZZ$.
Of these, the $\tau^+ \tau^-$ final state clearly yields no baryons, while the
numbers of baryons yielded by the final states $W^+ W^-$, $hZ$ and $ZZ$
are all independent of the annihilation centre-of-mass energy $2 m_\chi$. Only the
$b {\bar b}$ and $t {\bar t}$ final states yield numbers of baryons that increase
with the annihilation centre-of-mass energy.

\begin{figure}[htb!]
\begin{center}
\epsfig{file=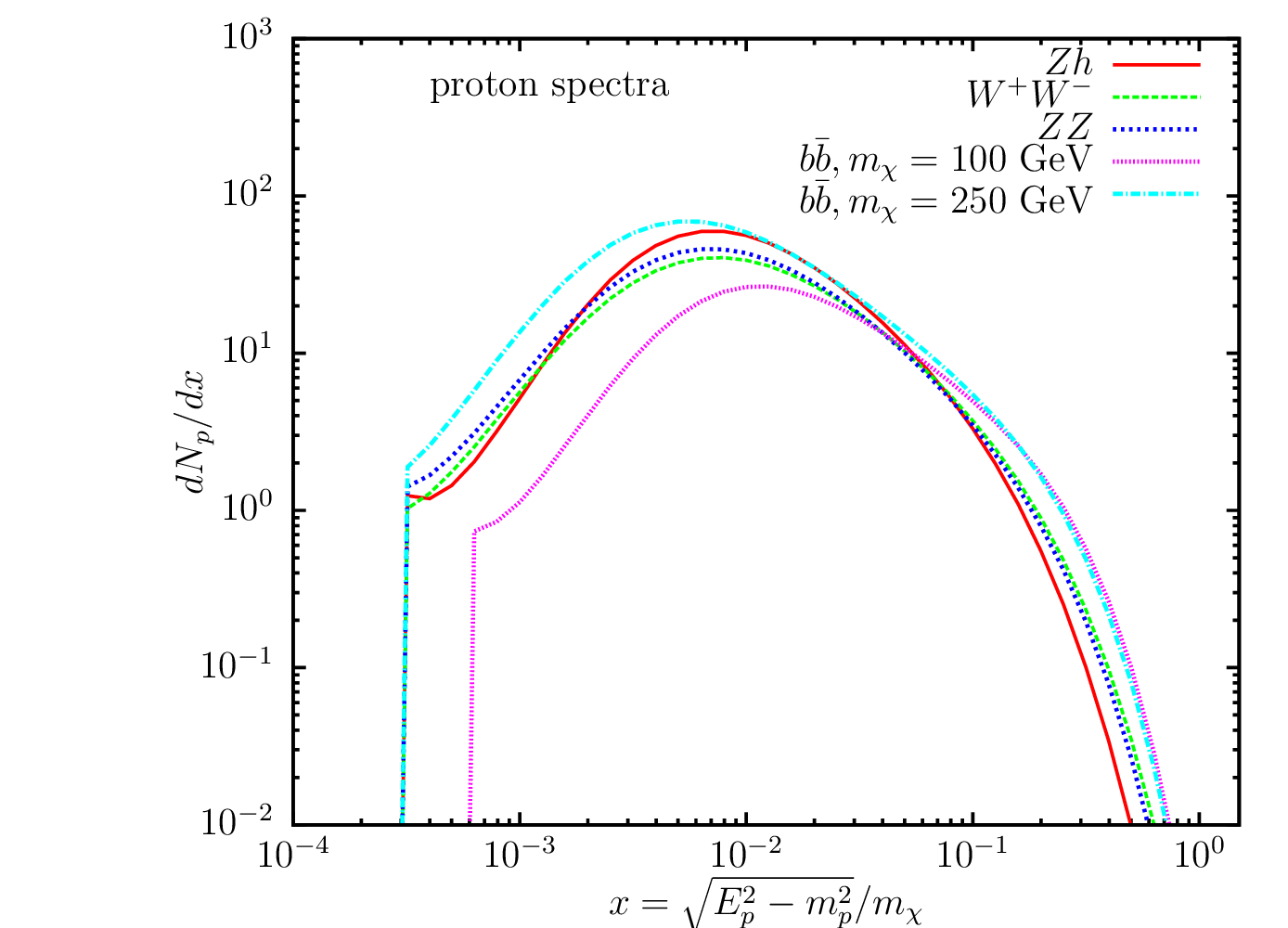,width=0.64\textwidth}  \\[1cm]
\epsfig{file=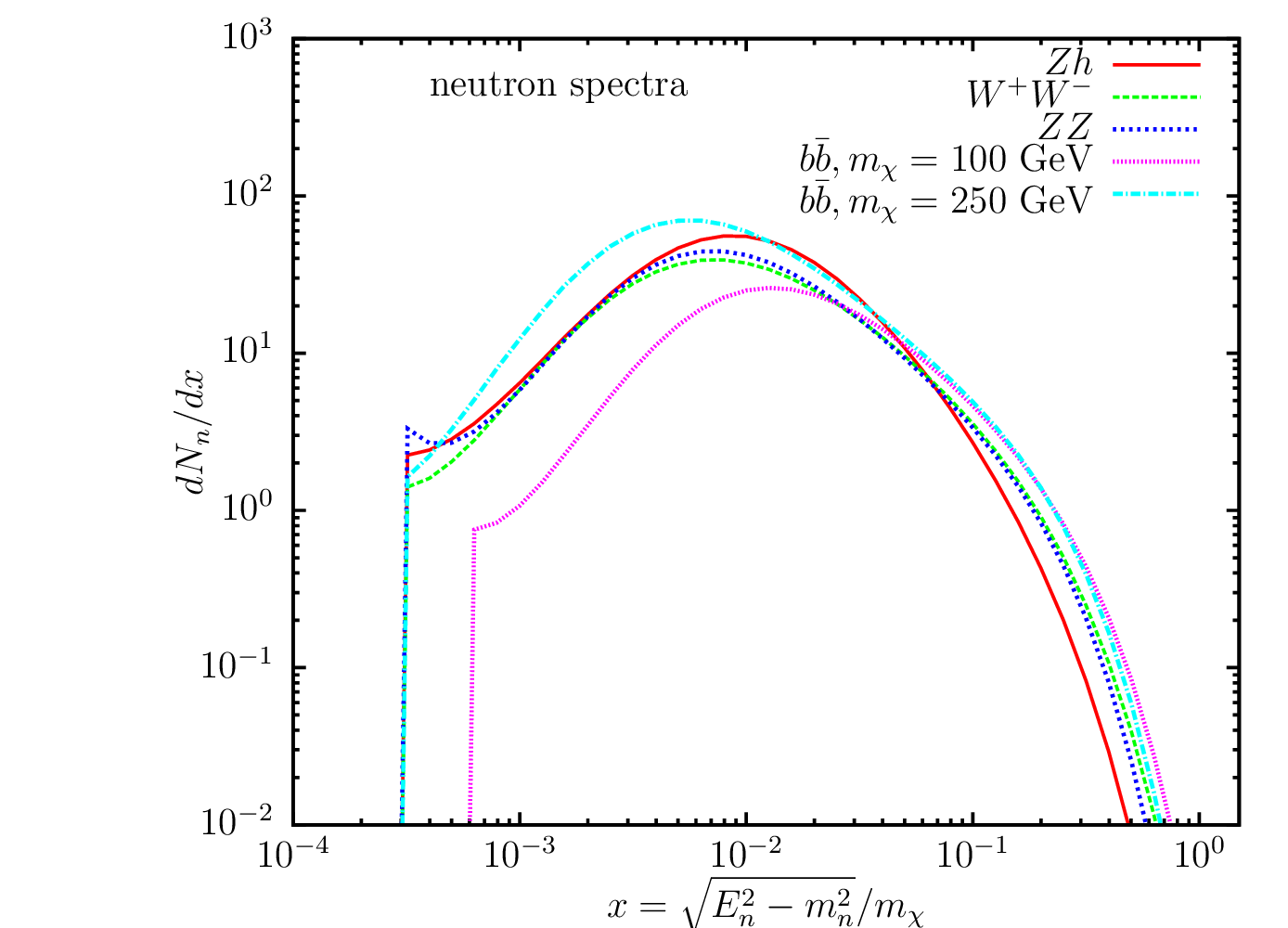,width=0.64\textwidth}
\caption{
\it The spectra of protons (upper panel) and neutrons (lower panel) injected by $\chi \chi$ annihilations
into the  $Z h$, $W^+ W^-$,  $ZZ$  and $b \bar{b}$ (for $m_\chi$=100 and $250$ GeV)
final states, as calculated using {\tt PYTHIA}.
}
\label{fig:baryons}
\end{center}
\end{figure}

Fig.~\ref{fig:baryons} displays the spectra of protons (upper panel) and neutrons
(lower panel) for the $W^+ W^-$, $hZ$ and $ZZ$ final states for $m_\chi = 250$~GeV, and for the
$b {\bar b} $ final state for $ m_\chi = 100, 250$~GeV, all calculated using
{\tt PYTHIA}. 
We display the number of protons or neutrons per unit of   the parameter
$x \equiv \sqrt{E_i^2-m_i^2}/m_\chi$, where $i=p,n$.
The proton and neutron spectra are almost identical. They differ  in the small $x$ region primarily because of 
the difference of $m_p$ and $m_n$.   We also see that the 
$W^+ W^-$, $hZ$ and $ZZ$ final states yield rather similar spectra, with
the spectrum from the $hZ$ final state rising slightly higher. The spectra
of baryons from $b {\bar b}$ final states rise from being lower at
$ m_\chi = 100$~GeV to being higher at $ m_\chi = 250$~GeV.

Together with Fig.~3 of~\cite{EOSgamma}, Fig.~\ref{fig:baryons} enables us to
understand the salient features of the baryon production rates shown in the
right panel of Fig.~\ref{fig:ratefom}. The large branching fraction for $\tau^+ \tau^-$
suppresses baryon injection along the coannihilation strip for $\tan \beta = 10$,
particularly for small $m_{1/2}$ but less so for large $m_{1/2}$ where the $W^+ W^-$
branching fraction grows.

\begin{figure}[htb!]
\begin{center}
\epsfig{file=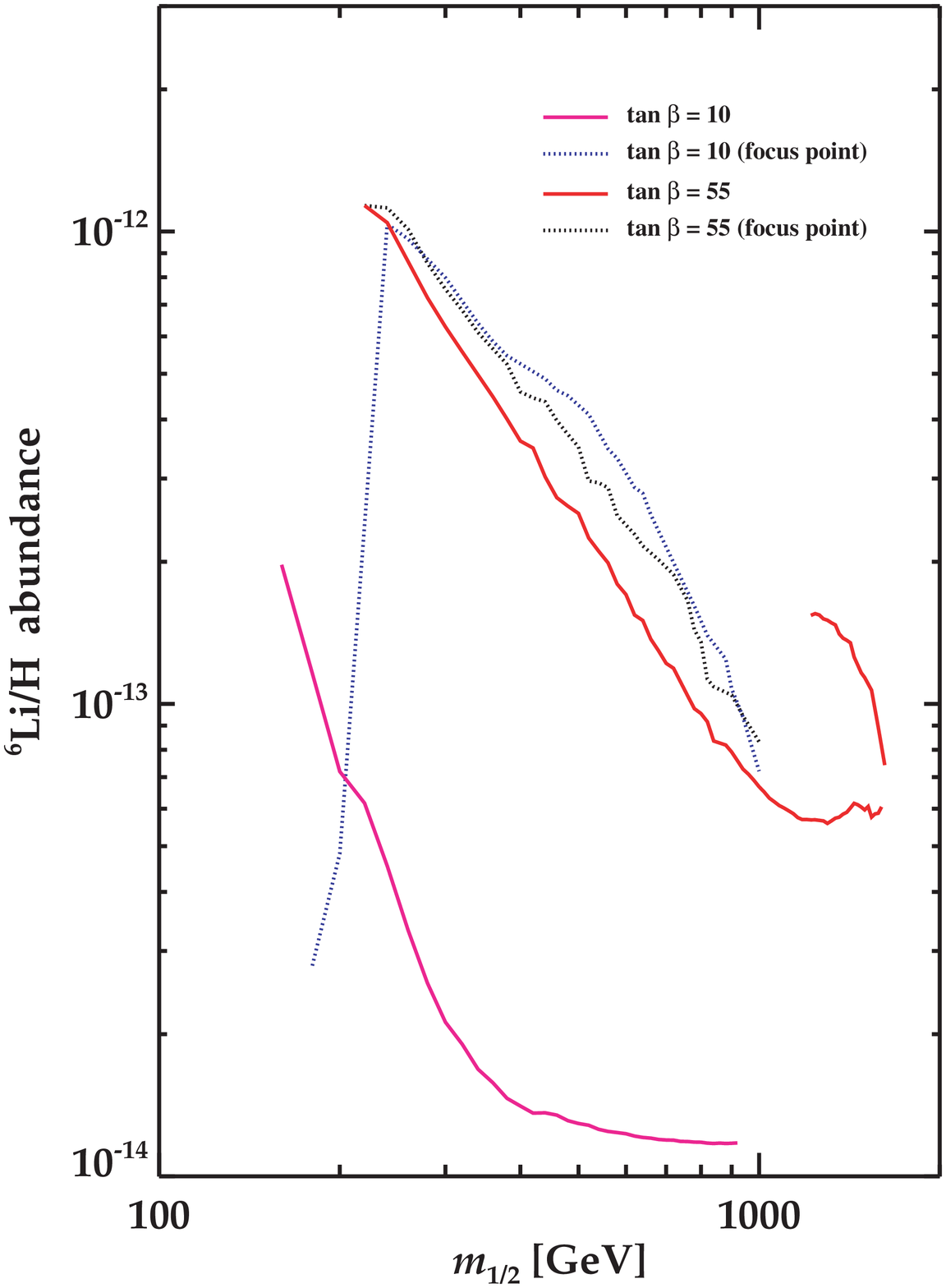,width=0.75\textwidth}
\caption{
\it The enhancement of the cosmological \li6 abundance as a function of
$m_{1/2}$ along the WMAP strips discussed in the text. The standard homogeneous
BBN value $\sim 10^{-14}$ is attained at large $m_{1/2}$ along the WMAP
coannihilation strip for $\tan \beta = 10$.}
\label{fig:li6results}
\end{center}
\end{figure}

Following their injection into the primordial plasma, some of the nucleons
cause spallation of \he4, yielding $A = 3$ nuclei as discussed above. These
are produced with large, nonthermal energies and subsequently thermalize,
but may previously induce $t (\alpha, n)\li6$ or $\he3 ( \alpha, p) \li6$ reactions.
Fig.~\ref{fig:li6results} displays the enhancement of the cosmological \li6 abundance
that we find along the WMAP strips discussed above. The homogeneous BBN
value $\sim 10^{-14}$ is attained at large $m_{1/2}$ along the WMAP
coannihilation strip for $\tan \beta = 10$, but much larger values are possible
along the other WMAP strips~\footnote{The region of enhanced \li6 along the WMAP coannihilation strip for $\tan \beta = 10$ with $m_{1/2} < 400$~GeV
is now excluded by the unsuccessful LHC searches for supersymmetry~\cite{LHC,CMS}.}, where
we find \li6/H $\sim 10^{-13}$ at large $m_{1/2}$ to $10^{-12}$ at small $m_{1/2}$.
We have found that the enhancement of \li6 scales very closely with the combination
$B_N \sigva / m_\chi^2$, as was to be expected.

We recall that the enhancement of \li6/H that would be required for consistency
with (\ref{Li6Li7}) is by a factor $\sim 1000$, rather than the factor of up to
$\sim 100$ that we find here. However, 
as we have noted there remains a question as to whether
or not the plateau ratio of 0.05 should be attributed to \li6. 
The abundance of \li6 we find here is potentially observable
and would in fact be seen as a plateau extending to low metallicities.
Optimistically, we could envision \li6 observations playing a role in
discerning between supersymmetric models.  In any case, 
we regard the enhancement we find as
already an interesting contribution to the analysis of the \li6 problem.

\subsection{Exploration of Non-Universal Higgs Models}

It is quite possible that some modifications of the CMSSM
might yield even greater enhancements of the \li6 abundance.
To be successful in this respect, it is apparent from (\ref{FoM}) that
such a model would require a relatively large annihilation
cross section $\langle \sigma v \rangle_{\rm ann}$ combined with a small value of $m_\chi$, as in the
focus-point region of the CMSSM. There, the relatively large value of $\langle \sigma v \rangle_{\rm ann}$
is made possible by the admixture of a Higgsino component in the $\chi$, and along this strip the low value
of $m_\chi$ is consistent with the LHC and other constraints~\cite{LHC}.

In an initial probe of other possibilities for a large enhancement of the \li6
abundance, we have explored the NUHM1 model, in which the
soft supersymmetry-breaking contributions to the Higgs masses have a
common value that differs from $m_0$. It is known that in this model the
Higgsino component in the LSP $\chi$ may be enhanced
at values of $m_{1/2}$ and $m_0$ away from the focus-point region,
thanks to a level-crossing transition at particular values of $\mu/m_{1/2}$ \cite{nuhm12}. In the CMSSM, 
the value of $\mu$ is generally fixed by applying the conditions for a consistent
electroweak vacuum. However, in the NUHM1 the value of $\mu/m_{1/2}$ can be adjusted by
varying the degree of non-universality in the soft supersymmetry-breaking Higgs
masses, enabling a WMAP-compatible relic density to be found in models with values of $(m_{1/2}, m_0)$
different from those allowed in the CMSSM.

We have explored the conditions under which such transition regions in the NUHM1 may yield an
enhancement of \li6/H similar to, or (possibly) greater than the value $\sim 10^{-12}$ attainable in the
CMSSM in the focus-point region. To this end, we have studied over a dozen NUHM1 parameter
planes. In no case did we find enhancements of \li6/H significantly larger than in the CMSSM (and
this is also the case in some planes we explored in the NUHM2, in which both Higgs soft supersymmetry-breaking
masses are treated as free, non-universal parameters).

\begin{figure}[htb]
\begin{center}
\epsfig{file=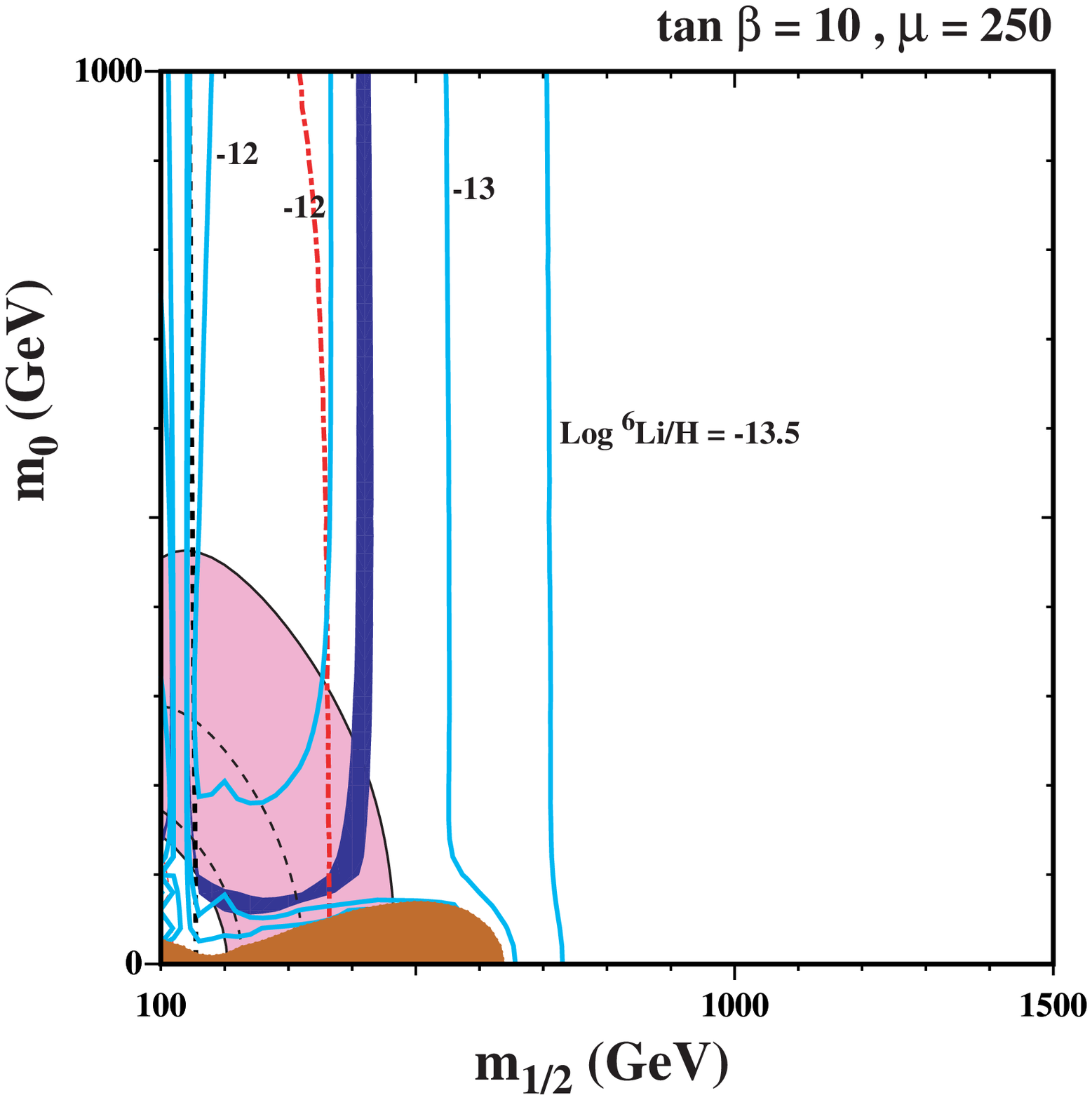,width=0.475\textwidth}
\epsfig{file=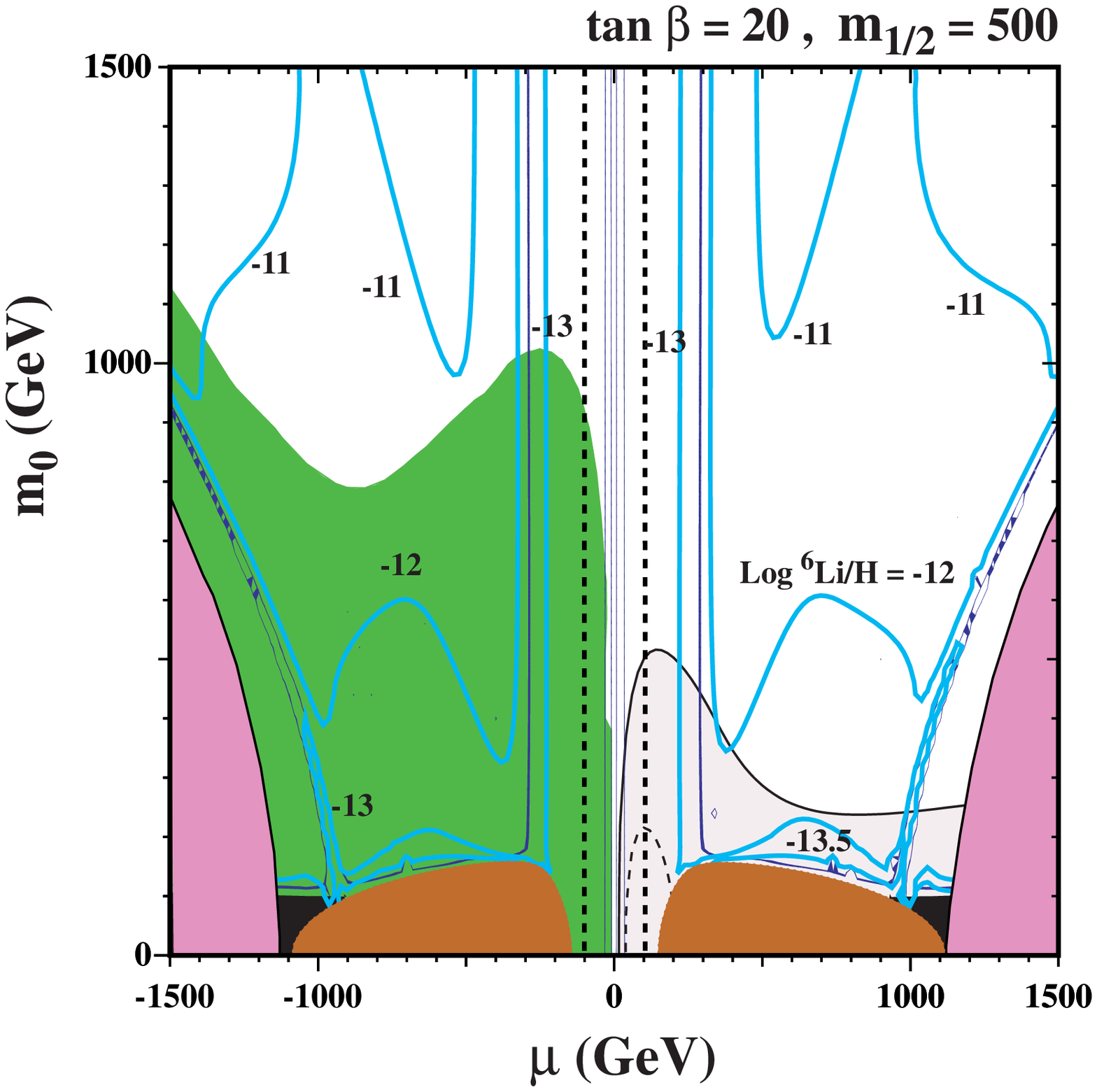,width=0.48\textwidth}
\caption{
\it Left: The NUHM1 $(m_{1/2}, m_0)$ plane for $\mu = 250$~GeV, $A_0 = 0$ and $\tan \beta = 10$,
and Right: the NUHM1 $(\mu, m_0)$ plane for $m_{1/2} = 500$~GeV, $A_0 = 0$ and $\tan \beta = 20$,
displaying contours of the \li6 abundance including the effects of late-time
$\chi \chi$ annihilations. Contours of the \li6 abundance
are coloured light blue, and the WMAP-compatible strips of
parameter space are shaded dark blue. The other shaded regions and lines have the same
meanings as in Fig.~\protect\ref{fig:planes}.}
\label{fig:NUHM1}
\end{center}
\end{figure}

Fig.~\ref{fig:NUHM1} shows results in a couple of
selected NUHM1 parameter planes. The left panel shows an $(m_{1/2}, m_0)$ plane for
$\tan \beta = 10, A_0 = 0$ and fixed $\mu = 250$~GeV. In this case, there is a near-vertical
WMAP-compatible strip in a transition region at $m_{1/2} \sim 400$~GeV.
This transition strip is compatible with the LEP Higgs constraint,
and the upper part of the strip above $m_0 \sim 700$~GeV is compatible with the constraints
imposed by LHC searches for sparticles. We see that the \li6/H ratio is remarkably
constant at $\sim 5 \times 10^{-13}$ along this strip. It would be possible to increase
\li6/H to $\sim 10^{-12}$ by choosing $\mu$ somewhat smaller, in which case the
WMAP-compatible strip would be at smaller $m_{1/2}$. In that case, the LHC
would enforce a stronger lower limit on $m_0$, closer to the CMSSM focus-point strip.
On the other hand, larger values of $\mu$ yield small values of the \li6 abundance, and we find
no increase in the \li6 abundance for larger $\tan \beta$.

The right panel of Fig.~\ref{fig:NUHM1} displays a $(\mu, m_{0})$ plane in the NUHM1 for
$\tan \beta = 20, A_0 = 0$ and fixed $m_{1/2} = 500$~GeV, at the lower end of the range allowed
by the LHC and other data for $m_0 < 1000$~GeV. In this case, 
there are near-vertical WMAP-compatible strips in transition region at $|\mu| \sim 300$~GeV, 
where \li6/H approaches $10^{-12}$. There is also a WMAP-compatible strip near
$\mu \sim 1000$~GeV that parallels the
region without a consistent electroweak vacuum (here caused by $m_A^2 < 0$), 
where \li6/H is again somewhat below
$10^{-12}$. Connecting these two regions is a co-annihilation segment at $m_0 \sim 100-200$~GeV where the \li6 abundance is relatively small. 
We have explored several other NUHM1 $(\mu, m_0)$ planes, finding that
increasing $m_{1/2}$ decreases the attainable value of \li6/H. We have also explored
several other projections of the NUHM1 and NUHM2, including $(m_A, m_{1/2})$, 
$(m_A, m_0)$, $(\mu, m_A)$ and $(m_1, m_2)$ planes, without finding values of \li6/H above $10^{-12}$.

\section{Summary and Conclusions}

We have demonstrated in this paper that in both the CMSSM and the NUHM1
it is possible that late neutralino LSP annihilations may enhance significantly
the cosmological \li6 abundance, without affecting significantly the BBN
abundances of the other light element Deuterium, \he3, \he4 and \li7~\cite{Jedamzik:2004ip}.
This enhancement may be up to two orders of magnitude, yielding \li6/H $\sim 10^{-12}$
compared to the BBN value $\sim 10^{-14}$.

As we have shown, this enhancement occurs typically when the neutralino LSP
is relatively light and has a large annihilation cross section, as occurs when the LSP
contains a strong Higgsino admixture. This phenomenon appears, in particular,
in the focus-point region of the CMSSM and in transition regions of the NUHM1.

While interesting for the debates on the astrophysical Lithium abundances, 
this enhancement falls short of resolving by itself the cosmological \li6 problem.
Further work could include a more exhaustive study of other supersymmetric
models, to see whether they could reconcile a larger enhancement with the
available theoretical, phenomenological, experimental and cosmological
constraints. Alternatively, is it possible that the height of the \li6 plateau may
receive contributions from other sources such as an early generation of stars,
or might the height of the \li6 plateau be over-estimated, perhaps because of
convective processes involving \li7 \cite{cayrel}? It is clearly
desirable to pin down more definitively the magnitude of the \li6 problem
by establishing more solidly the existence and height of the inferred \li6
plateau in halo stars. However, it already seems that a substantial
enhancement of the standard homogeneous BBN prediction for \li6/H
might be an interesting signature of supersymmetric models.

Finally, our work illustrates in detail 
the more general point that \li6 production should 
play a role in--and thus probe--{\em any} WIMP
dark matter scenario involving hadronic annihilation products \cite{karsten}.
Specifically, we have seen that \li6 production
is essentially guaranteed provided there are 
nonthermal nucleons injected with kinetic energies $\ga {\rm few}$ MeV.
We also find that the level of \li6 abundance due to
residual annihilations is model-dependent,
in our 
case spanning a range from 100 times the standard yield
down to an unobservable perturbation to this level.
The lessons for WIMP modelers would seem to be that
\li6 observations already provide important constraints
which one must test against, and that a confirmed detection
of primordial \li6--particularly if it is above the standard level--will
likely shed light on the details of the nonstandard physics which
produced it.

\section*{Acknowledgments}

We would like to thank R.V. Wagoner for interesting discussions on this problem.
The work of J.E. was supported partly by the London
Centre for Terauniverse Studies (LCTS), using funding from the European
Research Council 
via the Advanced Investigator Grant 267352. This Grant also supported visits by K.A.O. and 
V.C.S. to the CERN TH Division, which they thank for its hospitality. The work of F.L. and K.A.O. was supported in part
by DOE grant DE--FG02--94ER--40823 at the University of Minnesota.
The work of V.C.S. was supported by Marie Curie International
  Reintegration grant SUSYDM-PHEN, MIRG-CT-2007-203189.

\end{document}